\documentclass[pre,superscriptaddress,showkeys,showpacs,twocolumn]{revtex4-2}
\usepackage{graphicx}
\usepackage{latexsym}
\usepackage{amsmath}
\begin {document}
\title {Mean-field approximation and phase transitions in an Ising-voter model on directed regular random graphs}
\author{Adam Lipowski}
\affiliation{Faculty of Physics and Astronomy, Adam Mickiewicz University, Pozna\'{n} 61-614, Poland}
\author{Ant\'onio  Luis Ferreira}
\affiliation{Departamento de F\'{i}sica, I3N, Universidade de Aveiro,  Aveiro 3810-193, Portugal}
\author{Dorota Lipowska}
\affiliation{Faculty of Modern Languages and Literature, Adam Mickiewicz University, Pozna\'{n} 61-874, Poland}
\author{Aleksandra Napiera\l a-Batygolska}
\affiliation{Faculty of Physics and Astronomy, Adam Mickiewicz University, Pozna\'{n} 61-614, Poland}
%%%%%%%%%%%%%%%%%%%%%%%%%%%%%%%%%%%%%%%%%%%%%%%%%%%%%%%%%%%%%%%%%%%%%%%%%%%%%
\begin {abstract}
It is known that on directed graphs, the correlations between neighbours of a given site vanish and thus simple mean-field-like arguments can be used to describe exactly the behaviour of Ising-like systems. We analyse heterogeneous modifications of such models where a fraction of agents is driven by the voter or the antivoter dynamics.  It turns out that voter agents do not affect the dynamics of the model  and it behaves like a pure Ising model. Antivoter agents have a stronger impact since they act as a kind of noise, which weakens a ferromagnetic ordering.  Only when Ising spins are driven by the heat-bath dynamics, the behaviour of the model is correctly described by the mean-field approximation. The 
Metropolis dynamics generates some additional correlations that render the mean-field approach approximate. 
Simulations on annealed networks agree with the mean-field approximation but for the model with antivoters and with the Metropolis dynamics only its heterogeneous version provides such an agreement. Calculation of the Binder cumulant confirms that critical points in our models with the heat-bath dynamics belong to the Ising mean-field universality class. For the Metropolis dynamics, the phase transition is most likely discontinuous, at least for not too many antivoters. 

\end{abstract}
%\pacs{} 
%	\keywords{migration, Naming Game, language formation}

\maketitle
\section{Introduction}
Evolution of dynamical systems on heterogenous networks was examined in a variety of contexts such as, for example, disease spreading~\cite{pastor2001}, opinion formation~\cite{loreto}, neural activity~\cite{honey2009}, or fluctuations of financial markets~\cite{may2011}. Very often these complex systems can be to some extent described by a collection of interacting agents, whose states are represented by certain discrete, very often binary, variables. To analyse such simplified systems, effcient statistical-mechanics methods were developed and consequently various percolation problems~\cite{newman2010}, and Ising models~\cite{dorog2002,leone2002}, or epidemic spreading models~\cite{parshani2010} were examined on networks of different topologies. For many years, the description of such models as well as the study of their evolution and the analysis of possible phase transitions and critical points have stimulated numerous  researchers~\cite{gleeson,gleeson2011,mezard2009}.

Complexity of a society, of a collection of neurons in a brain, or of finanical markets is to some extent a consequence of strong heterogeneity of agents that consitute such systems and their accurate description should consider this factor. In some respects, such a heterogeneity is taken into account by the very structure of the network of interactions between agents. Indeed, such networks are very often strongly heterogeneous, for example in terms of the degree distribution of their vertices, a prime example of which are scale-free networks~\cite{barabasi2002}.

In the present paper, we examine models with dynamical heterogeneity, where agents are equipped with different dynamics. We use only very simple ones, namely the Ising~\cite{stauffer,ising2017}, the voter~\cite{redner2019}, or its less known modification, i.e., the antivoter~\cite{matlof,donnelly,huber}) dynamics.  Agents with such dynamics, partly due to their simplicity, were used in a multitude of studies but their mixture has not been studied that intensively~\cite{liplipfer2017,liplip2022,batlip}.

Interactions between our agents are modelled by directed random graphs. As a result, our models fall into the class of nonequilibrium systems with nonreciprocal interactions~\cite{fruchart,sanchez2002,lip2015}. Such systems have been recently studied intensively in various contexts ranging from inhibitatory and excitatory neurons~\cite{hatano}, to conformist and contrarian members of social groups~\cite{hong}, to oscillatory behaviour in spin models~\cite{bertin,avni,raajev}.

Directed networks have been already used for the analysis of statistical-mechanics systems. It is known that on such networks, the correlations between neighbours of a given site vanish and the dynamics of Ising-like systems substantially simplifies~\cite{derrida87,mezard88}. Consequently, closed-form expressions were obtained using the generating functional analysis for the steady-state characteristics in such systems~\cite{mimura}. Perhaps not surprisingly, expressions for, e.g., magnetization obtained using the generating functional analysis, which are expected to be exact, are identical to the simple mean-field approximation (MFA)~\cite{lipferlip2023}. However, the MFA reproduces the exact stead-state magnetization only when Ising spins are driven by the heat-bath dynamics. For the model with the Metropolis dynamics, applicability of techniques such as generating functional analysis is an open question and the exact solution could be very difficult to obtain. Comparison with numerical simulations shows that  the MFA description is only approximate~\cite{lipferlip2023} and the overall behaviour of the model is different and the phase transition becomes discontinuous.  In the present paper, we examine the suitability of the MFA and its heterogeneous version (HMFA) for  describing heterogeneous Ising/voter (or antivoter) systems.  We also analyse phase transitions and critical behaviour in such systems.

\section{Model}
We examine models with $N$ binary variables $s_i=\pm 1$, called agents or spins, placed on vertices~($i$) of a directed random graph.  To simplify the MFA description and eliminate effects related to the node degrees of variables, we restrict our analysis to regular graphs. The graphs are generated with a straightforward algorithm, which randomly selects $z$~neighbours (i.e., out-links) for each vertex (excluding connections to itself and multiple connections). As a result, we obtain a directed random graph, each vertex of  which has $z$~out-links. The number of in-links of a vertex has the Poisson distribution with the average value also equal to~$z$. The structure of such graphs is kept fixed during the evolution of the model and we call them quenched. Part of our simulations were also performed for annealed networks, in which neighbouring spins were selected anew upon each update of a given agent. Simulations were made for $z=8$, which is considerably larger than the percolation threshold $z=1$ for directed random graphs~\cite{dorog2001,newman2001}, and for such a choice most of the vertices belong to the giant cluster. On the other hand, for such a value of~$z$ and large~$N$, the  generated networks remain sparse graphs.

Agents in our model use either Ising-like or voter-like dynamics. One of these two types of the dynamics is assigned to each agent with probabilities~$p$ and~$1-p$, respectively. The assigned type of the dynamics (Ising or voter) is kept fixed. For Ising-like agents,  we examine separately the heat-bath and the Metropolis dynamics~\cite{newman-barkema}. 
In the version with the heat-bath dynamics, one randomly selects an agent, say~$s_i$, and if it is Ising-like, it is set  to~$+1$ with probability 
\begin{equation}
r(s_i\!=\!1)=\frac{1}{1+\exp(-2h_i/T)}, \ \  h_i=\sum_{j_i} s_{j_i},
\label{heat-bath}
\end{equation}
and  to~$-1$  with probability $1 - r(s_i\!\!=\!\!1)$.
The temperature-like parameter~$T$ controls the noise of the system and the summation in Eq.~(\ref{heat-bath}) is over the out-neighbours of the agent~$s_i$. 
We will also examine the case when Ising-like agents are driven by the Metropolis dynamics. In this dynamics, if a randomly selected agent~$s_i$ is Ising-like, it is flipped with probability $\min{[1,\exp(-2s_ih_i/T)]}$, where $h_i$~is defined in Eq.~(\ref{heat-bath}).

For voter-like agents, we examine (separately) two possibilities: (i)~voter agent, which takes the value of its randomly chosen out-neighbour, or (ii)~antivoter agent, which takes the opposite value of its randomly chosen out-neighbour.

As it follows from the above description, we analyse 4~types of the dynamics of agents in our models: (i)~heat-bath + voter, (ii)~heat-bath + antivoter (iii)~Metropolis + voter, and (iv)~Metropolis + antivoter. Considering that the random graphs on which these agents are placed are either quenched or annealed, this means that we actually examine 8~models.

The dynamics implemented for Ising-like agents suggests a similarity to the equilibrium Ising models but as we have already mentioned in the Introduction, the directedness of  random graphs as well as the presence of voter-like agents render our model nonequilibrium~\cite{sanchez2002,lip2015}.

Using numerical simulations, we calculated the time average of the magnetization $m=\frac{1}{N}\sum_{i=1}^N s_i$. Moreover, to get some insight into the critical behaviour and phase transitions in our models, we also calculated the Binder cumulant~\cite{binder1981} defined as 
\begin{equation}
U=1-\frac{<m^4>}{3<m^2>^2}.
\label{binder-def}
\end{equation}
%%%%%%%%%%%%%%%%%%%%%%%%%%%%%%%%%%%%%%%%%
\section{Results}

%%%%%%%%%%%%%%%%%%%%%%%%%%%%%
\subsection{Heat-bath dynamics}
\label{section-hb-dynamics}
%%%%%%%%%%%%%%%%%%%%%%%%%%%%%
The Ising model on directed random graphs with the heat-bath dynamics and in the absence of voter agents has been already examined~\cite{hatchett,mimura,lipferlip2023}. 
Using mean-field reasoning and assuming that the out-neighbours of a given site are uncorrelated, one can expect that in the stationary state, the probability $P_+$ that a randomly chosen spin equals~$+1$ satisfies the following equation~\cite{lipferlip2023}
\begin{equation}
P_+=\sum_{k=0}^z \frac{R_{z,k}}{1+\exp{[-4(k-z/2)/T]}}
\label{h-b-MFA}
\end{equation}
where  $R_{z,k}=\binom{z}{k}P_+^k (1-P_+)^{z-k}$. 
In Eq.~(\ref{h-b-MFA}), we assume that for each neighbour the probability that it is in the state~$+1$ is also equal to~$P_+$ (homogeneity).
The above nonlinear equation can be easily solved numerically, and knowing $P_+$, one can calculate the average magnetization as $m=2P_+-1$.

The absence of correlations between neighbours of a given site on directed networks was noticed by Derrida~\mbox{{\it et al.}} and it was used to solve some models of neural networks~\cite{derrida86,derrida87}.  Such a property also simplifies the so-called generating functional analysis~\cite{hatchett,mimura} and compact solutions were obtained for a class of Ising models (subject to a minor modification, namely the models are driven by parallel dynamics). One can easily notice that the MFA as formulated by Eq.~(\ref{h-b-MFA}) is exactly equivalent to the steady-state solutions as obtained using generating functional analysis. Precise Monte Carlo simulations~\cite{lipferlip2023} are also in very good agreement with predictions of Eq.~(\ref{h-b-MFA}).  It shows that very simple arguments lead us to  Eq.~(\ref{h-b-MFA}), which yields a remarkably successful description of these spin models. In the rest of this chapter, we will examine to what extent such a simple approach can describe a wider class of models.
%%%%%%%%%%%%%%%%%%%%%%%%%%%%%%%%%
\subsubsection{voter agents}
\label{section-hb-voter}
The simplest modification of the MFA that takes into account voter-like agents is straightforward. Namely, in the stationary state one expects that the model is still described by the single parameter $P_+$, which satisfies the following equation
\begin{equation}
P_+=p\sum_{k=0}^z \frac{R_{z,k}}{1+\exp{[-4(k-z/2)/T]}} +(1-p)P_+.
\label{h-b-voter}
\end{equation}
In the above equation, we assume that with probability~$p$, the chosen agent is Ising-like and evolves according to the heat-bath dynamics. With probability $1-p$, the selected agent is voter-like and  the probability that it will be set to~$+1$ is equal to $P_+$, because each out-neigbour is set to~$+1$ with probability~$P_+$.  

Let us notice, however, that the MFA for the mixed model (Eq.~(\ref{h-b-voter})) for any $p>0$ is equivalent to the MFA for the pure ($p=1$) Ising model (Eq.~(\ref{h-b-MFA})). This is a somewhat surprising feature, which shows that voter-like agents are in a sense irrelevant and even a very small concentration of Ising-like agents is sufficient to induce the ferromagnetic behaviour with magnetization exactly as in the pure Ising-like system. A similar behaviour was reported for the mixed Ising-voter model on complete graphs~\cite{liplip2022}.

\begin{figure}
\includegraphics[width=\columnwidth]{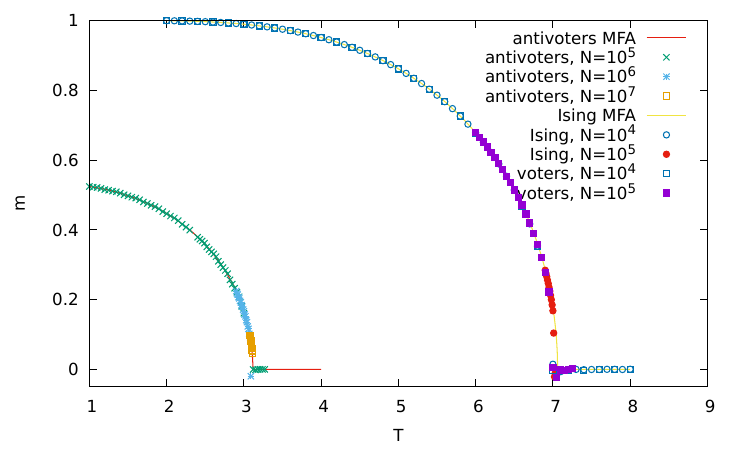}
\vspace{-8mm}

\caption{(Color online) The temperature dependence of the magnetization~$m$ for models with the heat-bath dynamics. Simulations were made for the pure ($p=1$) Ising model, for the Ising model with voters ($p=0.75$), and the Ising model with antivoters ($p=0.75$).  Numerical data demonstrate that  for the heat bath-dynamics, these models  are perfectly described by the single-site MFA (Eqs.~\ref{h-b-voter}) and~(\ref{h-b-antivoter}). Moreover, the presence of voters does not affect steady-state characteristics, such as the magnetization~$m$, and the solutions for $p=1$ and $p<1$ are the same.}
\label{mag-heat-bath}

\end{figure}

Monte Carlo simulations do support such a claim. In Fig.~\ref{mag-heat-bath}, we present the values of magnetization~$m$ as a function of temperature~$T$. One can notice that the results for the pure Ising model ($p=1$) are in very good agreement with the mixed model with $p=0.75$.

%%%%%%%%%%%%%%%%%%%%%%%%%%%%%%%%%%
%%%%%%%%%%%%%%%%%%%%%%%%%%%%%%%

As we have already mentioned, for the pure Ising model with the heat-bath dynamics on directed random graphs, the MFA is expected to provide the exact description of its steady state~\cite{lipferlip2023}. Actually, it was numerically demonstrated that the MFA seems to be exact for a larger class of models whose dynamics resemble the heat-bath dynamics, namely, where the configuration of surrounding spins determines the probability that a given spin will be set to a certain state. Since the dynamics of voter and (examined in the next subsection) antivoter agents also belong to this class, we can expect that the MFA for the mixed Ising-voter model (Eq.~(\ref{h-b-voter})) will also give the exact values of the steady-state~$P_+$ and the magnetization~$m$. 

\begin{figure}
\includegraphics[width=\columnwidth]{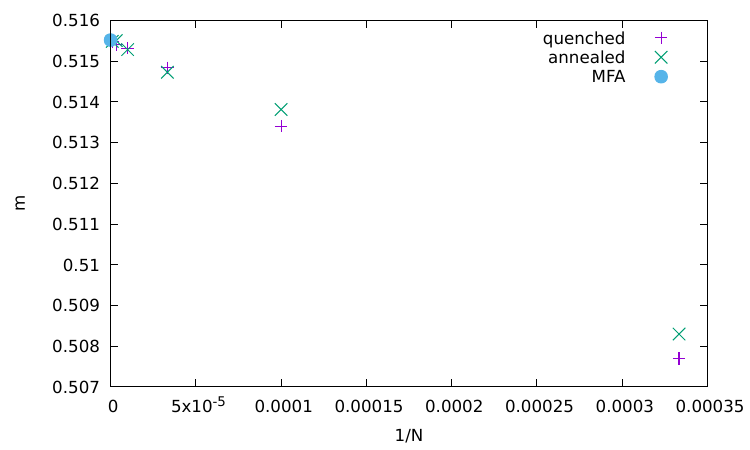}
\caption{(Color online)  The magnetization~$m$ as a function of~$1/N$ for the Ising model with voters and with concentration of the Ising spins $p=0.75$. Calculations were made for the model with the heat-bath dynamics at $T=6.5$. For the increasing system size~$N$, for both quenched and annealed distributions of links, good agreement  with the MFA prediction $m=0.515518$ (bullet) can be seen. The results presented are averages over 100 independent runs (with new distributions of links and Ising/voters agents) and statistical errors are smaller than the size of the plotted symbols.} 
\label{finite-size-voter}
\end{figure}

To verify such a possibility, we made a more detailed analysis of the results of simulations for the mixed model with $p=0.75$ and at $T=6.5$. Numerical results, presented in Fig.~\ref{finite-size-voter}, show that in the limit $N\rightarrow \infty$, the steady-state magnetization~$m$ indeed converges to the value obtained from the solution of Eq.~(\ref{h-b-voter}). Extrapolating, we obtain that in the limit $N\rightarrow \infty$, Monte Carlo simulations agree with the MFA with a relative accuracy of~$\sim 10^{-4}$. 

We made simulations for both quenched and annealed graphs, and in both cases very good agreement with the MFA was obtained. The agreement of the MFA with simulations for annealed networks is generally expected~\cite{pastor2015}.
What is not entirely obvious is the fact that in the presence of voters, our model is still correctly described by the single parameter~$P_+$. In principle, one could expect that Ising and voter agents should be described by different probabilities. We will return to this point in section~\ref{sub-metrop-antivoters}.

%%%%%%%%%%%%%%%%%%%%%%%%%%%%%%%%%%
%%%%%%%%%%%%%%%%%%%%%%%%%%%%%%%

\begin{figure}
\includegraphics[width=\columnwidth]{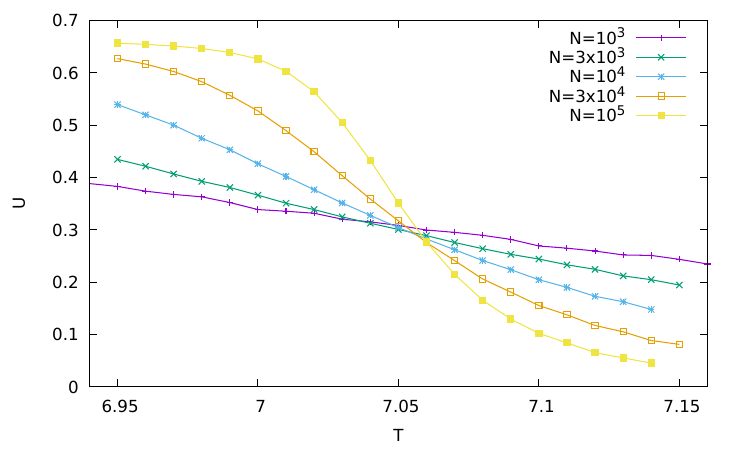}
\vspace{-8mm}

\caption{(Color online) The temperature dependence of  the Binder cumulant~$U$ for the pure ($p=1$) Ising model with the heat-bath dynamics. From the numerical solution of the MFA (Eq.~(\ref{h-b-MFA})), we obtain that in this case $T_c=7.064$, which is in good agreement with the crossing point of the cumulant values. At the critical temperature, we estimate the value of the Binder cumulant as~0.28(1). Let us notice that at the critical point of the mean-field (equilibrium) Ising models, the Binder cumulant is expected to be equal to~0.2705~\cite{brezin}. }
\label{binder-hb-ising}

\end{figure}

\begin{figure}
\includegraphics[width=\columnwidth]{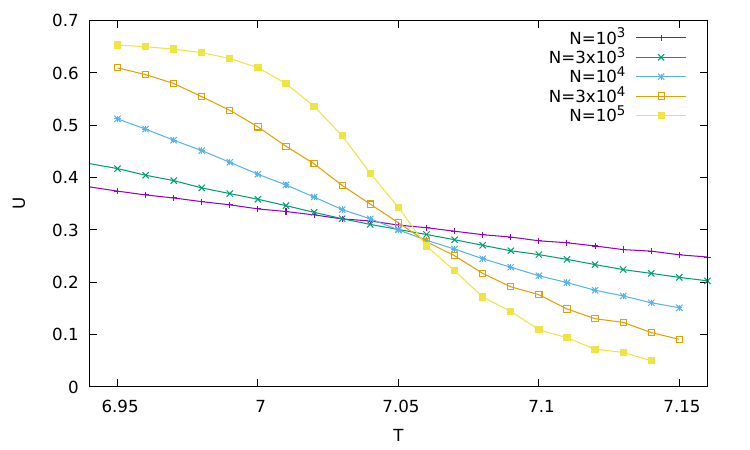}
\vspace{-8mm}

\caption{(Color online) The temperature dependence of  the Binder cumulant~$U$ for the Ising model with voters ($p=0.75$) and driven by the heat-bath dynamics. From the numerical solution of the MFA (Eq.~(\ref{h-b-voter})), we obtain that also in this case $T_c=7.064$, which is in good agreement with the crossing point of the cumulant values. At the critical temperature, we estimate the value of the Binder cumulant as~0.28(1).}
\label{binder-hb-voter}

\end{figure}

Since the mean-field approximation describes the steady-state properties of our Ising-like models, we expect that their critical behaviour also belongs to the so-called Ising mean-field universality class. To classify a model to a given universality class, one usually calculates critical exponents that describe the behaviour of some quantities at or in the vicinity of the critical point. However, keeping in mind that our models are nonequilibrium and hence lack the definition of the free energy and canonical distribution, it is not entirely  clear how to even define some of these critical exponents, perhaps except for the exponent~$\beta$ that describes the decay of the order parameter~$m$ at criticality. A very important parameter used to locate and classify the critical point is the Binder cumulant~$U$. Numerical calculations for the pure Ising and mixed ($p=0.75$) Ising/voter models are presented in Fig.~\ref{binder-hb-ising} and Fig.~\ref{binder-hb-voter}, respectively.
They show that in both cases the crossing points for various system sizes~$N$ are at nearly the same temperature, which is, moreover, very close to the  value $T=7.064$ as obtained from the solution of the MFA (for $z=8$). Moreover, we estimate the value of the Binder cumulant at the crossing point as $0.28(1)$ and such a value is very close to~0.2705 expected for the Ising mean-field universality class~\cite{brezin}.
%%%%%%%%%%%%%%%%%%%%%%%%%%%%%%%%%%
%%%%%%%%%%%%%%%%%%%%%%%%%%%%%%%

%%%%%%%%%%%%%%%%%%%%%%%%%%%%%%%%%%
%%%%%%%%%%%%%%%%%%%%%%%%%%%%%%%

\subsubsection{antivoter agents}
In this subsection, we present results for the mixed Ising/antivoter model, in which we distribute Ising and antivoter agents with the concentration~$p$ and~$1-p$, respectively. In the case when an antivoter agent is selected, it takes the opposite value  of its randomly chosen neighbour. Such behaviour of antivoter agents frustrates the system and weakens the ferromagnetic ordering. 

For the mixed Ising/antivoter model,  the modification of the MFA takes the following form
\begin{equation}
P_+=p\sum_{k=0}^z \frac{R_{z,k}}{1+\exp{[-4(k-z/2)/T]}} +(1-p)(1-P_+).
\label{h-b-antivoter}
\end{equation}

Contrary to the MFA for the Ising/voter model (Eq.~(\ref{h-b-voter})), the above equation is not equivalent to the MFA for the pure Ising model (Eq.~(\ref{h-b-MFA})).
An exemplary solution of Eq.~(\ref{h-b-antivoter}) for $p=0.75$ shows that, indeed, the  magnetization and critical temperature are much lower than for the Ising and Ising/voter models (Fig.~\ref{mag-heat-bath}). Monte Carlo simulations are in very good agreement with the solution obtained using Eq.~(\ref{h-b-antivoter}).
Actually, the argumentation about uncorrelated neighbours on random directed graphs is also applicable to the Ising/antivoter model and we expect that the solution of Eq.~(\ref{h-b-antivoter}), similarly to the Ising and Ising/voter models, should give the exact values of the steady state~$P_+$ and magnetization~$m$.

\begin{figure}
\includegraphics[width=\columnwidth]{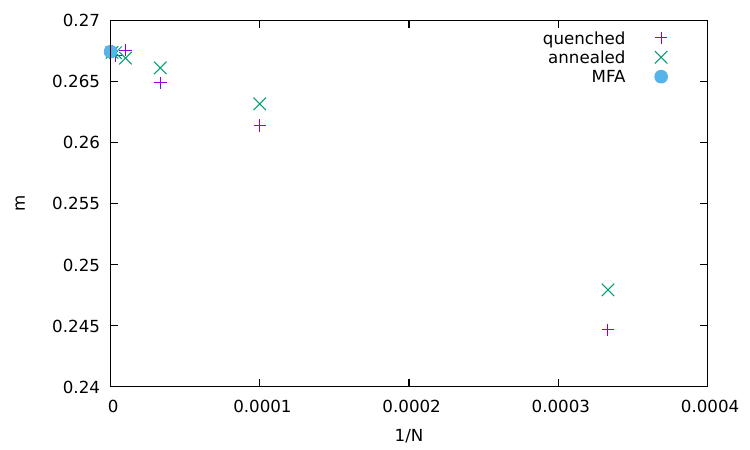}
\caption{(Color online)  The magnetization~$m$ as a function of~$1/N$ for the Ising model with antivoters and with concentration of Ising spins $p=0.75$. Calculations were made for the model with the heat-bath dynamics at $T=2.8$. For the increasing system size~$N$, for both quenched and annealed distributions of links, good agreement  with  the MFA prediction $m=0.267438$ (bullet) can be seen. The results presented are averages over 100 independent runs and statistical errors are smaller than the size of the plotted symbols.} 
\label{finite-size-antivoter}
\end{figure}

Numerical simulations support such an expectation. In Fig.~\ref{finite-size-antivoter} we present the magnetization calculated for the Ising/antivoter model with $p=0.75$ and $T=2.8$. One can notice that in the limit $N\rightarrow\infty$, the results clearly converge to the value obtained from the solution of Eq.~(\ref{h-b-antivoter}). Such a convergence is seen for both quenched and annealed distributions of Ising and antivoter agents.

%%%%%%%%%%%%%%%%%%%%%%%%%%%%%%%%%%
%%%%%%%%%%%%%%%%%%%%%%%%%%%%%%%

\begin{figure}
\includegraphics[width=\columnwidth]{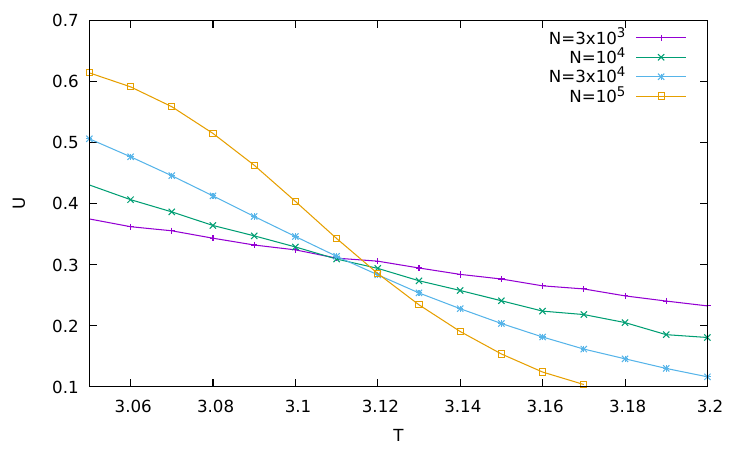}
\vspace{-8mm}

\caption{(Color online) The temperature dependence of  the Binder cumulant~$U$ for the Ising model with antivoters \mbox{($p=0.75$)} and driven by the heat-bath dynamics. The MFA (Eq.~\ref{h-b-antivoter}) predicts that in this case $T_c=3.117$, which is in good agreement with the crossing point of the cumulant values. At the critical temperature, we estimate the value of the Binder cumulant as~0.28(1). }
\label{binder-hb-antivoter}

\end{figure}

We also calculated the Binder cumulant for the Ising/antivoter model with $p=0.75$. Numerical results show (Fig.~\ref{binder-hb-antivoter}) that the crossing temperature is in very good agreement with the MFA (Eq.~(\ref{h-b-antivoter})). We estimate the value of the Binder cumulant at the crossing point as~0.28(1), which is also in good agreement with the value expected for the Ising mean-field universality class~\cite{brezin}.

%%%%%%%%%%%%%%%%%%%%%%%%%%%%%%%%%%
%%%%%%%%%%%%%%%%%%%%%%%%%%%%%%%

\subsection{Metropolis dynamics}
For simulations of equilibrium systems, one can use various kinds of dynamics and the heat bath dynamics is one of them. These dynamics are defined in such a way that they all reproduce the canonical distribution of a given equilibrium system. Hence, equilibrium characteristics obtained from simulations with different dynamics are the same. As we have  already mentioned, the Ising model on directed networks is a nonequilibrium model and different dynamics do not necessarily give the same results. Indeed, we have recently  shown that on directed random graphs, Ising models with the heat-bath and the Metropolis dynamics have different behaviour~\cite{lipferlip2023}.  In particular, with the heat-bath dynamics, the model has a continuous phase transition, while with the Metropolis dynamics, it seems to exhibit a discontinuous one. 

A discontinuous transition is also predicted by the MFA.  Such an approximation for the model with the Metropolis dynamics  can be obtained from the requirement that in the steady-state, transitions $+1 \rightarrow -1$ and $-1 \rightarrow +1$ take place with the same rate. Although a more formal derivation for a broader class of systems can be obtained from the stationary solution of the corresponding Master equation~\cite{gleeson},  the resulting steady-state equation for our model has the form~\cite{lipferlip2023}
\begin{widetext}
\begin{equation}
P_+\left[ \sum_{k=0}^{z/2} R_{z,k}+\sum_{k=z/2+1}^{z} R_{z,k} \exp{(2z-4k)/T} \right] = (1-P_+)\left[ \sum_{k=0}^{z/2-1} R_{z,k}\exp{(4k-2z)/T}+\sum_{k=z/2}^{z} R_{z,k}  \right]
\label{metrop-Ising-MFA}
\end{equation}
\end{widetext}

\begin{figure}
\includegraphics[width=\columnwidth]{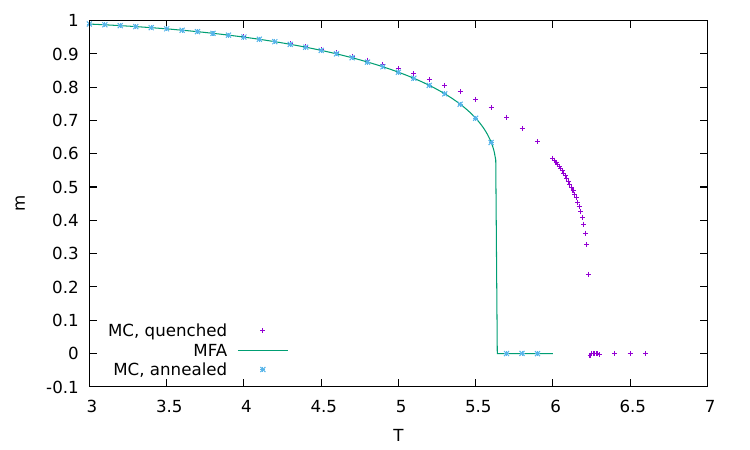}
\caption{(Color online)  The magnetization~$m$ as a function of the temperature~$T$ for the Ising ($p=1$) model with the Metropolis dynamics. Monte Carlo simulations  were made for $N=10^5$ but close to the transition points, $N=10^6$ was used. For the annealed networks,  perfect agreement with the MFA (Eq.~(\ref{metrop-Ising-MFA})) is obtained.} 
\label{mag-metrop-pav0}
\end{figure}

Numerical simulations for a quenched network deviate from the predictions of Eq.~(\ref{metrop-Ising-MFA}) at higher temperatures (Fig.~\ref{mag-metrop-pav0}). Such a deviation does not appear for the heat-bath dynamics and it is a consequence of different rules of the Metropolis dynamics.
Let us notice that in Eq.~(\ref{metrop-Ising-MFA}), we assume that the spin under consideration is independent of the neighbouring spins (hence the product form of the left and right sides of Eq.~(\ref{metrop-Ising-MFA}). Validity of such an assumption is usually limited and the above approach is only approximate.  
Consequently, contrary to the model with the heat-bath dynamics, Eq.~(\ref{metrop-Ising-MFA}) does not provide the exact description of the steady-state behaviour.
Such a deficiency is a well-known feature of the MFA and one can develop more sophisticated versions that to some extent improve this approximation~\cite{gleeson}. We also did simulations for an annealed network and, as expected,  the results are in very good agreement with the MFA (Fig.~\ref{mag-metrop-pav0}). 

\begin{figure}
\includegraphics[width=\columnwidth]{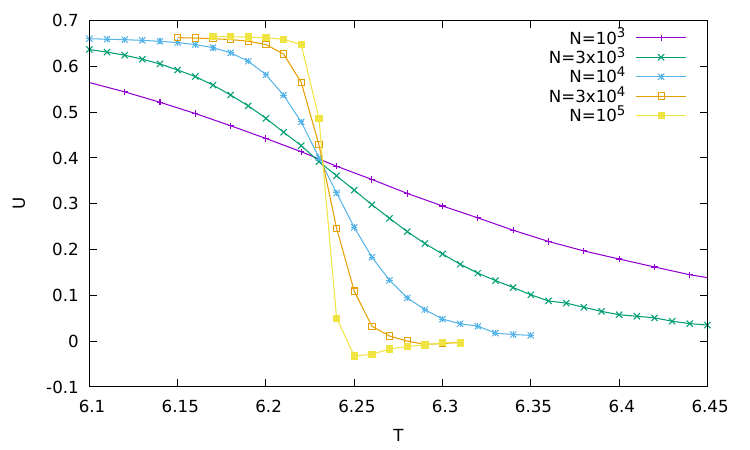}
\vspace{-8mm}
\caption{(Color online) The temperature dependence of  the Binder cumulant~$U$ for the Ising model with the Metropolis dynamics. The rather large value of~$U$ at the crossing point and the drop to the negative values for the largest~$N$ suggest a discontinuous transition~\cite{vollmayr}.}
\label{binder-metrop-pav0}

\end{figure}

From the behaviour of the order parameter, as obtained from Monte Carlo simulations, it is rather difficult to determine the type of the phase transition. More reliable predictions can be made by calculating the Binder cumulant~$U$. Numerical results show (Fig.~\ref{binder-metrop-pav0}) that at the crossing point, $U$ has a large value~$\sim 0.4$, which is considerably larger than~0.2705, which one could expect at the continuous Ising-like transitions~\cite{brezin}. Moreover, for the largest sytem size $N=10^5$, we observe that just above the transition, $U$ drops to negative values. It has already been argued that such a behaviour of the Binder cumulant can be typical of discontinuous transitions~\cite{vollmayr}. Thus, the calculation of the Binder cumulant gives further support for the claim that the Ising model on directed random graphs with the Metropolis dynamics exhibits a discontinuous transition.

%%%%%%%%%%%%%%%%%%%%%%%%%%%%%%%%%%
%%%%%%%%%%%%%%%%%%%%%%%%%%%%%%%

%%%%%%%%%%%%%%%%%%%%%%%%%%%%%%%%%%
%%%%%%%%%%%%%%%%%%%%%%%%%%%%%%%
In the following subsections, we would like to examine how voter and antivoter agents affect the behaviour of the Ising model with the Metropolis dynamics.

%%%%%%%%%%%%%%%%%%%%%%%%%%%%%%%%%%%%%%%%%
\subsubsection{voter agents}
Analysing in the previous section the model with the heat-bath dynamics, we have already mentioned that voter agents have no effect on the stationary properties of the model. For the model with the Metropolis dynamics, we observe a similar effect and the simplest explanation refers to the MFA. In the presence of Ising voters, the steady-state generalization of Eq.~(\ref{metrop-Ising-MFA}) takes the form
\begin{widetext}
\begin{multline}
pP_+\left[ \sum_{k=0}^{z/2} R_{z,k}+\sum_{k=z/2+1}^{z} R_{z,k} \exp{(2z-4k)/T} \right] +(1-p)P_+(1-P_+)= \\ p (1-P_+)\left[ \sum_{k=0}^{z/2-1} R_{z,k}\exp{(4k-2z)/T}+\sum_{k=z/2}^{z} R_{z,k}  \right] +(1-p)(1-P_+)P_+
\label{metrop-voter-MFA}
\end{multline}
\end{widetext}
The term $(1-p)P_+(1-P_+)$ on the left hand side corresponds to a randomly chosen agent which is voter ($1-p$) in the state~$+1$ ($P_+$) with its randomly chosen neighbour being in the state~$-1$ ($1-P_+$). An analogous expression appears on the right hand side but it corresponds to the randomly chosen agent which is voter in the state~$-1$ with its randomly chosen neighbour being in the state~$+1$. Let us notice that these terms are exactly the same. They cancel out and the resulting equation is equivalent to Eq.~(\ref{metrop-Ising-MFA}). Such a behaviour is similar to the  model with the heat-bath dynamics (Section \ref{section-hb-dynamics}), where the MFA for the model with voter agents (Eq.~(\ref{h-b-voter})) is also shown to be equivalent to the MFA for the pure Ising model (Eq.~(\ref{h-b-MFA})).

\begin{figure}
\includegraphics[width=\columnwidth]{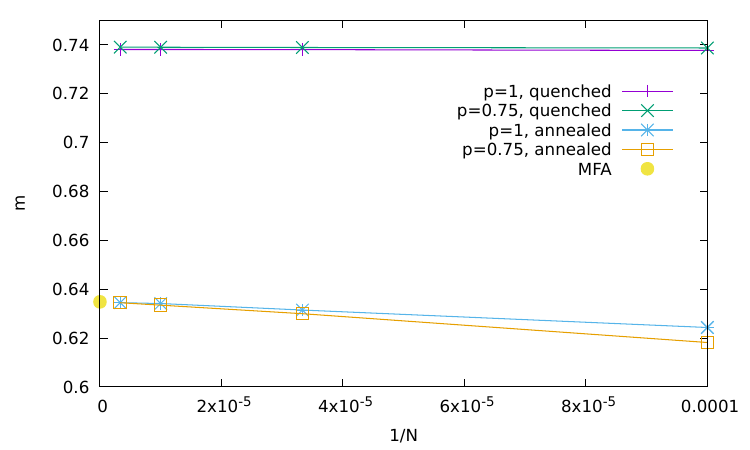}
\caption{(Color online)  The magnetization~$m$ as a function of~$1/N$ for the Ising model with the Metropolis dynamics at $T=5.6$. 
For annealed interactions, the results for  $p=1$ and $p=0.75$ seem  to be asymptotically (for $N\rightarrow\infty$)  the same  and in perfect agreement with the MFA (Eq.~(\ref{metrop-Ising-MFA})). Results for quenched interactions differ from the MFA. Although not visible, the results in the presence of voters ($p=0.75$) are slightly different from those for the pure Ising ($p=1$).} 
\label{metrop-anneal-size}
\end{figure}

We confirmed numerically that, indeed, voter agents have no influence on the stationary properties of the model (Fig.~\ref{metrop-anneal-size}). For models on annealed networks, such independence is expected since in this case the mean-field arguments should be exact. Less obvious is independence for quenched networks, but as shown in Fig.~\ref{metrop-anneal-size}, it also holds in this case, eventhough the values of the magnetization are much different from the MFA.
%%%%%%%%%%%%%%%%%%%%%%%%%%%%%%%%%%
%%%%%%%%%%%%%%%%%%%%%%%%%%%%%%%

\subsubsection{antivoter agents}\label{sub-metrop-antivoters}
In the presence of anti-voters, the MFA can be written as
\begin{widetext}
\begin{multline}
pP_+\left[ \sum_{k=0}^{z/2} R_{z,k}+\sum_{k=z/2+1}^{z} R_{z,k} \exp{(2z-4k)/T} \right] +(1-p)P_+^2= \\ p (1-P_+)\left[ \sum_{k=0}^{z/2-1} R_{z,k}\exp{(4k-2z)/T}+\sum_{k=z/2}^{z} R_{z,k}  \right] +(1-p)(1-P_+)^2
\label{metrop-antivoter-MFA}
\end{multline}
\end{widetext}

\begin{figure}
\includegraphics[width=\columnwidth]{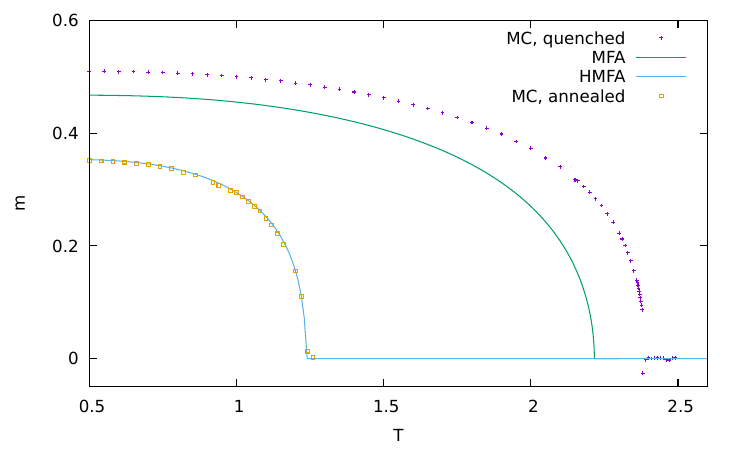}
\caption{(Color online)  The magnetization~$m$ as a function of the temperature~$T$ for the Ising model with antivoters and with the Metropolis dynamics for $p=0.75$. Calculations were made for $N=10^5$ but close to the transition points, $N=10^6$ was used.} 
\label{mag-metrop-pav025}
\end{figure}

\begin{figure}
\includegraphics[width=\columnwidth]{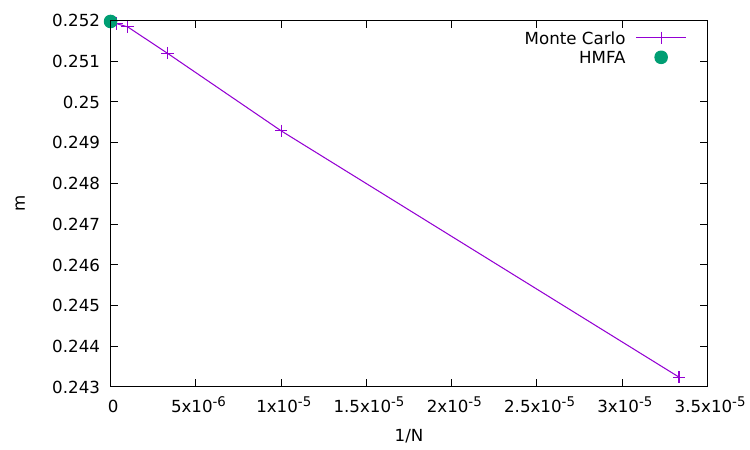}
\caption{(Color online)  The magnetization~$m$ as a function of~$1/N$ for the Ising model with antivoters ($p=0.75$) driven by the Metropolis dynamics at $T=1.1$ on the annealed networks. 
In the limit $N\rightarrow\infty$, the  results are in perfect agreement with the HMFA ($m=0.5219..$).} 
\label{t11-anneal-size}
\end{figure}

The numerical solution of the above equation and Monte Carlo simulations show that the presence of antivoters reduces the magnetization of the model (Fig.~\ref{mag-metrop-pav025}). One can notice, however, that agreement of the MFA (Eq.~(\ref{metrop-antivoter-MFA})) with simulations for quenched and even for annealed networks is rather poor.
As we will demonstrate, this is because such a simple version of the MFA is based on the single quantity~$P_+$ and neglects the heterogeneity of the system, namely, the fact that there are some Ising agents and some antivoter ones. To take into account such a heterogeneity, we have to introduce different parameters~$P_+^I$ and~$P_+^A$, which denote the probabilities that a randomly chosen agent being of the Ising or antivoter type, respectively,  is in the state~$+1$. In such a case, in the stationary state, we require that for both types of agents, the transitions $+1 \rightarrow -1$ and  $-1 \rightarrow +1$ take place with the same rate.
Such a heterogeneous mean-field approximation (HMFA) is thus formulated as
\begin{widetext}
\begin{align}
&P_+^I\left[ \sum_{k=0}^{z/2} R_{z,k}+\sum_{k=z/2+1}^{z} R_{z,k} \exp{(2z-4k)/T} \right] = (1-P_+^I)\left[ \sum_{k=0}^{z/2-1} R_{z,k}\exp{(4k-2z)/T}+\sum_{k=z/2}^{z} R_{z,k}  \right]  \label{metrop-antivoter-HMFA1}\\
&P_+^A=1-P_+ \label{metrop-antivoter-HMFA2}
\end{align}
\end{widetext}
where 
\begin{equation}
P_+=pP_+^I+(1-p)P_+^A. 
\label{def_P}
\end{equation}
The above set of equations can be easily solved numerically and the results show that the HMFA is in perfect agreement with the Ising model with antivoters on annealed networks (Fig.~\ref{mag-metrop-pav025}, Fig.~\ref{t11-anneal-size}).

Let us notice that for the HMFA for the Ising model with voters, Eq.~(\ref{metrop-antivoter-HMFA2}) takes the form $P_+^A=P_+$. Using Eq.~(\ref{def_P}), one obtains $P_+^A=P_+^I$, and Eq.~(\ref{metrop-antivoter-HMFA1}) becomes equivalent to Eq.~(\ref{metrop-Ising-MFA}), namely, to the MFA for the pure Ising model. It means that in the presence of voters, the model is heterogeneous at the microscopic level but it does not break its mean-field homegeneity.

Let us also notice that similar arguments show that the HMFA for the model with the heat-bath dynamics and in the presence of voters is also equivalent to the MFA for the pure Ising model (Eq.~(\ref{h-b-MFA})). Moreover, the HMFA for the model with the heat-bath dynamics and with antivoters is also equivalent to the MFA. Indeed, in this case the HMFA equations have the form
\begin{align}
&P_+^I=\sum_{k=0}^z \frac{R_{z,k}}{1+\exp{[-4(k-z/2)/T]}} \label{h-b-HMFA1} \\
&P_+^A=1-P_+ \label{h-b-HMFA2}
\end{align}
Using Eq.~(\ref{def_P}) and Eq.~(\ref{h-b-HMFA2}), we obtain $P_+^I=(2P_+-1+p-pP_+)/p$. With such an expression  for $P_+^I$, Eq.~(\ref{h-b-HMFA1}) can be easily shown to be equivalent to Eq.~(\ref{h-b-antivoter}), namely, to the MFA for the model with the heat-bath dynamics (and with antivoters).

%%%%%%%%%%%%%%%%%%%%%%%%%%%%%%%%%%
%%%%%%%%%%%%%%%%%%%%%%%%%%%%%%%

%%%%%%%%%%%%%%%%%%%%%%%%%%%%%%%%%%
%%%%%%%%%%%%%%%%%%%%%%%%%%%%%%%

\begin{figure}
\includegraphics[width=\columnwidth]{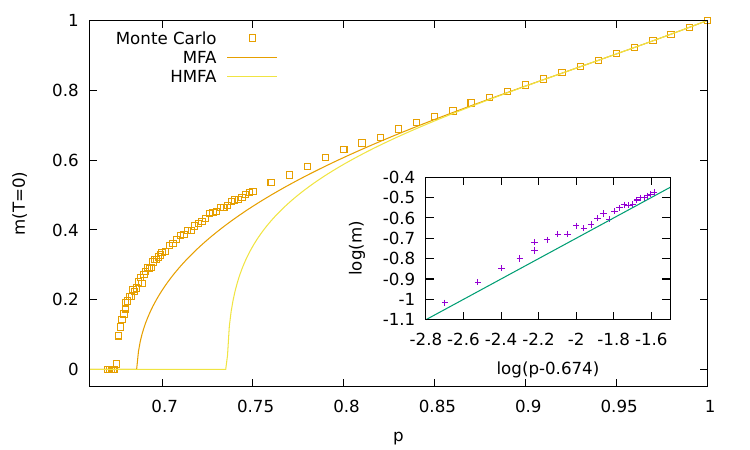}
\vspace{-0mm}
\caption{(Color online) The concentration dependence of the zero-temperature magnetization for the Ising model with antivoters and with the Metropolis dynamics on  quenched networks. The inset shows (decimal logarithm scale) that in the vicinity of the transition point, the magnetization seems to follow a power-law decay $m\sim (p-p_c)^{1/2}$ with $p_c=0.674$.}
\label{magnet-zero-temp}
\end{figure}

\begin{figure}
\includegraphics[width=\columnwidth]{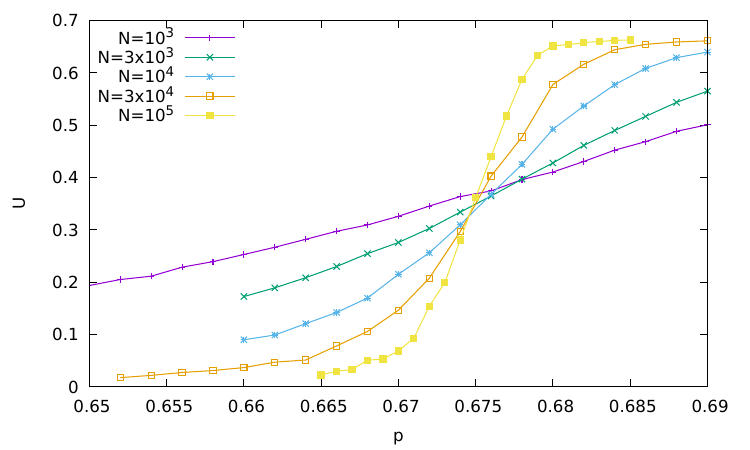}
\vspace{-0mm}
\caption{(Color online) The concentration dependence of  the Binder cumulant~$U$ for the Ising model with antivoters and with the Metropolis dynamics on quenched networks  at $T=0$.}
\label{bind-metrop-t0}
\end{figure}

Antivoters take the opposite orientation to neighbouring Ising spins so they can be considered as generating some kind of  noise.
Such a noise, when sufficiently strong,  can destroy a ferromagnetic ordering even in the absence of thermal noise. Indeed, we calculated the zero-temperature magnetization and when concentration of antivoters is sufficiently large (i.e., $p$ is sufficiently small), the ferromagnetic ordering is destroyed (Fig.~\ref{magnet-zero-temp}). Our simulations suggest that the model undergoes a continuous Ising-type phase  transition at $p=p_c=0.674$ with the expected power-law decay of the magnetization. Numerical calculation of the Binder cumulant supports such a scenario (Fig.~\ref{bind-metrop-t0}). In particular, at the crossing point $U\sim 0.30(4)$, and such value is also quite close to the Ising mean-field estimation 0.2705~\cite{brezin}.  Moreover, the drop of~$U$ to negative values is not observed even for the largest~$N$, which suggests a continuous transition in this case. Let us notice that the MFA (Eq.~(\ref{metrop-antivoter-MFA})) is in better agreement with the simulation data than the HMFA (Eq.~(\ref{metrop-antivoter-HMFA1})--Eq.~(\ref{metrop-antivoter-HMFA2})). As we demonstrated in our paper, the HMFA seems to be exact for models on annealed networks, but on quenched networks, simpler MFA turns out to be more accurate. Similar behaviour can be seen in Fig.~\ref{mag-metrop-pav025} for some temperature dependent characteristics.

%%%%%%%%%%%%%%%%%%%%%%%%%%%%%%%%%%
%%%%%%%%%%%%%%%%%%%%%%%%%%%%%%%

%%%%%%%%%%%%%%%%%%%%%%%%%%%%%%%%%%
%%%%%%%%%%%%%%%%%%%%%%%%%%%%%%%

\begin{figure}
\includegraphics[width=\columnwidth]{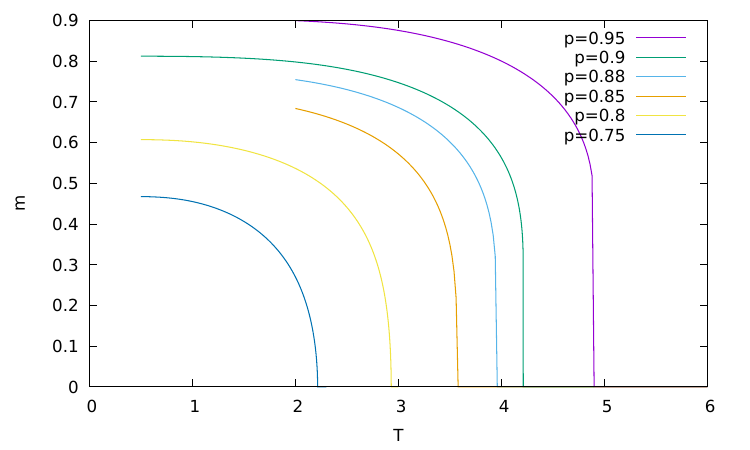}
\vspace{-0mm}
\caption{(Color online) The temperature~$T$ dependence of the magnetization~$m$ as obtained from the solution of the MFA (Eq.~(\ref{metrop-antivoter-MFA})) for the Ising model with antivoters and with the Metropolis dynamics.}
\label{metrop-analytical}
\end{figure}

We do not present the comparison with the Monte Carlo simulations but the solution of the  MFA (Eq.~(\ref{metrop-antivoter-MFA})) shows that the temperature of the discontinuous phase transition decreases for increasing concentration of antivoter agents (Fig~\ref{metrop-analytical}). Moreover, the transition turns into continuous one for $p<0.8$.

Simulations at $T=0$ (Fig.~\ref{magnet-zero-temp}-Fig.~\ref{bind-metrop-t0}) show that a sufficiently strong noise generated by antivoters leads to the critical behaviour and continuous phase transition.

As shown in Fig.~\ref{metrop-analytical}, critical fluctuations from antivoters could change the type of the phase transition in this model even when coupled with a thermal noise.

However, a reliable numerical verification of such a behaviour  for the model on quenched graphs, with a possible tricritical point separating continuous and discontinuous transitions, would be computationally very demanding and is left for the future.

%%%%%%%%%%%%%%%%%%%%%%%%%%%%%%%%%%
%%%%%%%%%%%%%%%%%%%%%%%%%%%%%%%

\section{Summary and Conclusions}

In our paper, we analyzed a class of stochastic models on regular random directed graphs in their quenched and annealed versions. Agents in our models use heterogenous dynamics---some of them (Ising-like) are driven by the heat-bath or Metropolis dynamics, and others use voter or antivoter rules. In addition to Monte Carlo simulations, we used mean-field approximations. Our intention was to examine the behaviour of such nonequilibrium and heterogeneous models but also to check the validity and accuracy of the mean-field description of their stationary characteristics.

Analysis of the results of simulations  shows that for models on quenched networks with the heat-bath dynamics and with extrapolation to infinitely large graphs, the mean-field approximation (MFA) most likely provides an exact  description even in the presence of voters or antivoters. We also show that for the heat-bath dynamics, the heterogeneous mean-field approximation (HMFA), which seems to take into account the dynamical heterogeneity of the model, is actually equivalent to the MFA.

For  models on quenched networks with the Metropolis dynamics, the MFA and HMFA provide only approximate descriptions. As we have already argued~\cite{lipferlip2023}, some discrepancies with simulations could be attributed to certain correlations between neighbouring spins, which appear for the Metropolis dynamics but are irrelevant for the heat-bath dynamics. For models on quenched networks (and with the Metropolis dynamics), the MFA seems to be more accurate than the HMFA. This is somewhat surprising taking into account the heterogeneity of the dynamics of our models.

%%%%%%%%%%%%%%%%%%%%%%%%%%%%%%%
\begin{table}
\begin{center}
\begin{tabular}{|c|c|c|c|}
\hline
\multicolumn{4}{|c|}{\textbf {Ising ($p$) + voter (1-$p$)}} \\
\hline
\multicolumn{2}{|c|}{\textbf {Heat-bath}} & \multicolumn{2}{|c|}{\textbf {Metropolis}}\\
\hline
\multicolumn{2}{|c|}{\textbf {quenched = annealed}} & \textbf {quenched  }& \textbf {annealed} \\
\hline
\multicolumn{4}{|c|}{for any $p>0$ equivalent to pure Ising ($p=1$)} \\
\hline
\multicolumn{2}{|c|}{MFA  exact} & MFA  approx. & MFA  exact \\
\hline
\multicolumn{4}{|c|}{HMFA is equivalent to MFA} \\
\hline
\multicolumn{2}{|c|}{continuous phase trans.} & \multicolumn{2}{|c|}{discontinuous phase trans.}\\
%\noalign{\vskip-2pt}.
\hline
\end{tabular}
\caption{Main results for the Ising-voter model.}
\end{center}
\end{table}
%%%%%%%%%%%%%%%%%%%%%%%%%%%%%%%

%%%%%%%%%%%%%%%%%%%%%%%%%%%%%%%
\begin{table}
\begin{center}
\begin{tabular}{|c|c|c|c|}
\hline
\multicolumn{4}{|c|}{\textbf {Ising ($p$) + antivoter (1-$p$)}} \\
\hline
\multicolumn{2}{|c|}{\textbf {Heat-bath}} & \multicolumn{2}{|c|}{\textbf {Metropolis}}\\
\hline
\multicolumn{2}{|c|}{\textbf {quenched = annealed}} & \textbf {quenched }& \textbf {annealed} \\
\hline
\multicolumn{2}{|c|}{HMFA=MFA exact} & HMFA$\neq$MFA  approx. & HMFA exact \\
\hline
\multicolumn{2}{|c|}{continuous phase trans.} & \multicolumn{2}{|c|}{discont. or continuous phase trans.}\\
\hline
\multicolumn{4}{|c|}{Increasing fraction of antivoters weakens ferromagnetic} \\
\multicolumn{4}{|c|}{ ordering. Continuous phase transition at $T=0$.} \\
\hline
\end{tabular}
\caption{Main results for the Ising-antivoter model.}
\end{center}
\end{table}
%%%%%%%%%%%%%%%%%%%%%%%%%%%%%%%

As expected, mean-field approximations reproduce correctly the behaviour of our models on annealed networks, as we numerically verified in some cases. Antivoters orient oppositely to neighbouring agents and apparently they generate stronger heterogeneity than voters. In the presence of antivoters, only the heterogeneous version of the mean-field approximation (HMFA) describes correctly the model on annealed networks with the Metropolis dynamics. It should be also emphasized that in the presence of voters, the HMFA is equivalent to the MFA (for both the heat-bath and the Metropolis dynamics). In other words, eventhough the dynamics of voters is much different than that of the Ising spins, they do not generate the mean-field heterogeneity.

Ising-like agents are driven by temperature-dependent dynamics and our models undergo ferromagnetic-paramagnetic phase transitions. Since the networks of interactions are random graphs, one could expect that these transitions belong to the Ising
 mean-field universality class. Calculation of the Binder cumulant to some extent   confirmed such expectations. Let us notice that arguments that the Binder cumulant remains universal at some critical points refer to the notion of the free energy and canonical distribution. Our numerical results suggest that such equilibrium statistical-mechanics methodology can be extended to some nonequilibrium systems.

For the pure Ising model with the Metropolis dynamics, calculation of the Binder cumulant suggests that the temperature-driven phase transition is discontinuous. The presence of antivoters generates some kind of noise, which, if sufficiently strong, induces a continuous transition, as confirmed by some zero-temperature simulations. This suggests that for a  sufficiently large fraction of antivoters, a temperature-driven  phase transition also becomes continuous perhaps with a tricritical point separating these different phase transitions. Such a behaviour can be seen in the MFA and it would be interesting to confirm it for the model on quenched networks. However, such calculations seem to be much more demanding and are left for the future.

In our paper we analysed models with different dynamics (heat bath, Metropolis), networks (quenched, annealed) and heterogeneities (voters, antivoters).
Brief summary of our main results is collected in Tables~I and~II.

Overall, our work demonstrates that for some dynamical systems with heterogeneous dynamics and placed on directed random graphs, mean-field approximations provide a reliable, and sometimes even exact, description.  For equilibrium systems, it is known that for the stationary characteristics, the choice of the spin dynamics is usually unimportant. As we have shown, for nonequilibrium models on directed networks, some fine dynamical details, such as the choice of the spin dynamics or the presence of dynamical heterogeneities, considerably change the behaviour of the system. It would certainly be desirable to better understand which factor is actually responsible for the type of phase transition in such systems. Directed, or at least nonsymmetric, networks of interactions seem to be an important component of neural networks, opinion-forming processes, or some oscillatory setups. Their dynamics are much more intricate than those of our Ising-(anti-)voter agents, and it would be very interesting to extend the analysis presented in our paper to such more complex yet more realistic systems.

%HMFA is known to be effective for network heterogeneity. We examine the role of heterogenous dynamics. 

% antivoter model:  P. Donnelly and D. Welsh. The antivoter problem: random 2-colourings of graphs. Graph Theory and Combinatorics, 133–144, 1984. MR777170

\section*{Acknowledgements} A.L.F. was supported by funding from the i3n-associated laboratory LA/P/0037/202, within the scope of the projects
UIDB/50025/2020 and UID-P/50025/2020, ﬁnanced by national funds through the FCT/MEC.

\bibliography{abbrev_titles,biblio}

\end {document}